# Monolithically Integrated Optical Convolutional Processors on Thin Film Lithium Niobate

*Ruixue Liu,[†] Rongbo Wu,[†,]\* Yong Zheng,[†] Yuan Ren, Boyang Nan, Min Wang\*, Yunpeng Song, and Ya Cheng\**

Ruixue Liu, Yuan Ren, Boyang Nan, Prof. Ya Cheng
State Key Laboratory of Precision Spectroscopy, East China Normal University, Shanghai 200062, China

Ruixue Liu, Rongbo Wu, Yong Zheng, Yuan Ren, Boyang Nan, Min Wang, Yunpeng Song, Prof. Ya Cheng
The Extreme Optoelectromechanics Laboratory (XXL), School of Physics and Electronic Science, East China Normal University, Shanghai 200241, China
E-mail: rbwu@phy.ecnu.edu.cn, mwang@phy.ecnu.edu.cn

Prof. Ya Cheng
Hefei National Laboratory, Hefei 230088, China
Shanghai Research Center for Quantum Sciences, Shanghai 201315, China
Collaborative Innovation Center of Extreme Optics, Shanxi University, Taiyuan 030006, China
E-mail: ya.cheng@siom.ac.cn

Ruixue Liu, Rongbo Wu, Yong Zheng Contributed equally to this work

Funding: National Natural Science Foundation of China (12192251, 12334014, 12134001, 12304418, 12474378). Innovation Program for Quantum Science and Technology (2021ZD0301403). Shanghai Municipal Science and Technology Major Project (2019SHZDZX01)

Keywords: optical convolutional processors, photonic integrated circuits, thin-film lithium niobate, optical true delay line

Photonic neural networks (PNNs) of sufficiently large physical dimensions and high operation accuracies are envisaged as an ideal candidate for breaking the major bottlenecks in the current artificial intelligence architectures in terms of latency, energy efficiency and computational power. To achieve this vision, it is of vital importance to scale up the PNNs and in the meantime reduce the high demand on the dimensions required by the PNNs. The underlying cause of this strategy is the enormous gap between the scales of photonic and electronic integrated circuits. Here, we demonstrate monolithically integrated optical convolutional processors on thin film lithium niobate (TFLN) to enable large-scale programmable convolution kernels and in turn greatly reduce the dimensions required by the subsequent fully connected layers. Experimental validation achieves high classification accuracies of 96%/86% on the MNIST/Fashion-MNIST datasets and 84.6% on the AG News dataset, while dramatically reducing the required subsequent fully connected layer dimensions to 196×10 (from 784×10) and 175×4 (from 800×4), respectively. Furthermore, our devices can be driven by commercial field-programmable gate array (FPGA) systems, a unique advantage in addition to their scalable channel number and kernel size, our architecture provides a solution to build practical machine learning photonic devices.

# 1.Introduction

Artificial intelligence has emerged as a transformative force reshaping socioeconomic structures, scientific paradigms, and human capabilities, yet its exponentially growing computational demands are creating sustainability barriers that impede its further progress. [1,2] Photons possess unique properties, including wide bandwidth, high parallelism, and extremely low power consumption for information processing. When laser beams carrying large amounts of data pass through specifically designed photonic devices, the data are processed following the physical laws that describe the propagation of the laser beam in the photonic devices. The principle establishes the foundation of photonic neural networks (PNNs). Recently, the continuous advancement of integrated photonics technology has led to the development of numerous PNNs based on various architectures across different material platforms. [3–17] These PNNs have demonstrated computational power that rivals that of state-of-the-art electronic devices while achieving significantly higher energy efficiency and lower latency by orders of magnitude. Despite significant achievements, the practical adoption of PNNs faces fundamental physical constraints. The primary limitation stems from the diffraction-limited feature sizes (>1μm) of photonic components, which impose significantly larger physical footprints compared to nanoscale electronic transistors. This size disparity directly translates to constrained matrix operation dimensions in implemented PNN systems. Notably, the largest experimentally demonstrated photonic matrix multiplier achieves merely 64 dimensions,[17] falling far short of the 784-dimensional operations required for even basic tasks like MNIST classification. This dimensional gap highlights a critical scalability challenge that hinders the practical deployment of PNNs in real-world applications.

Convolutional neural networks (CNNs) suggest viable pathways to address the fundamental scalability challenges currently limiting PNN deployment. At the heart of the CNN architecture, convolutional layers employ trainable filters that systematically traverse the input space, enabling efficient localized feature detection.[18] Together with pooling layers that perform spatial downsampling to condense feature representations, the convolutional layers effectively reduce the input dimensions of subsequent fully connected (FC) layers, resulting in significantly reduction in computational complexity and memory requirements. These advantages have made CNNs one of the predominant approaches for image processing and natural language processing.[19,20] However, the selection of convolutional kernel size critically influences layer performance and presents distinct trade-offs. Smaller kernels provide better computational efficiency, but their limited receptive fields inherently restrict their ability to capture long-range dependencies. While large kernels excel at processing complex patterns through expansive receptive fields, their substantial computational overhead from enlarged parameter spaces outweigh the benefits of the reduced computational cost in the subsequent FC layers for von Neumann architectures.

Optical convolutional processors (OCPs) offer a viable solution for this trade-off. Large convolutional operations can dramatically reduce the dimensionality of subsequent FC layers, and thus progressively bridge the critical gap between current PNN capabilities and the demanding requirements of practical applications. Meanwhile, OCPs can exploit the intrinsic advantages of optical computing to circumvent the exponential growth in power consumption that plagues electronic systems during kernel upscaling. Typically, OCPs can be realized through two fundamental architectural strategies: wavelength division multiplexing (WDM) systems,[21–25] which distribute data across the discrete spectral lines of optical frequency combs, and time division multiplexing (TDM) architectures, which allocate signals through optical true delay lines (OTDLs).[26,27] Early implementations of OCPs primarily utilized mature silicon-on-insulator (SOI) and silicon nitride platforms.[28–32] The recent development of high-quality thin-film lithium niobate (TFLN) wafers, along with the rapid advances in nanofabrication methodologies, has facilitated the realization of TFLN-based OCPs.[27,33] TFLN provides significant advantages owing to its intrinsically large electro-optic coefficient, which enables both high-speed operation and ultra-low-power phase modulation. Combined with remarkably low propagation loss characteristics (~3 dB/m),[34] these properties establish TFLN as a superior platform for constructing large-scale, high-performance and energy-efficient PICs for deep learning implementations. While significant progress has been made, it is still necessary to conquer several key challenges. First of all, kernel size scalability currently faces practical constraints, evidenced by the fact that most integrated implementations currently only demonstrate basic 3×3 kernels due to limitations in the number of independently tunable components. Secondly, architecture can be optimized as many systems employ fixed kernel (e.g., horizontal/vertical edge detectors) rather than fully trainable kernels. These non-optimized kernels, which are not explicitly trained for target datasets, inevitably increase computational overhead by necessitating larger input and hidden layer dimensions in subsequent FC layers. Thirdly, while benchtop arbitrary waveform generators (AWGs) with tens or even hundred Gbaud capabilities enable optimal utilization of PICs' high bandwidth for data encoding, their high cost and non-integrability pose fundamental barriers to practical implementation in compact systems. By contrast, integrated electronic platforms such as FPGAs, which are typically limited to ≤10 GS/s operation with sub-2 GHz bandwidths, demand both increased parallelism to maintain high throughput and longer low-loss OTDLs to accommodate increased sampling intervals. Lastly, monolithic integration of all critical photonic components remains a key technical requirement especially when facing the critical demands on increasing stability and scalability.

Here we present a monolithically integrated OCPs architecture that combines CMOS-voltage-drivable high-speed data encoding module, low-loss and precisely calibrated delay module, and multi-channel independently programable kernel weight tuning units on one platform. Based on this architecture, we demonstrate two OCPs: a 4-channel 4×4 kernel chip for image feature extraction and a single-channel 1×8 kernel variant for natural language processing. Both devices are fabricated using photolithography-assisted chemical mechanical etching (PLACE) technology, which enables ultra-low-loss waveguides fabrication and a large writing field.[10,34–37] The big convolutional kernels with arbitrarily programmable weights enable our devices to significantly reduce the FC layer dimensions while maintaining competitive accuracy: for MNIST and Fashion-MNIST datasets, the FC layer dimensions were reduced to 196×10 (from 784×10) while maintaining 96% and 86% classification accuracy, respectively; for the AG News dataset, an 84.6% accuracy was achieved with reduced FC dimensions of 175×4 (from 800×10). Our architecture features scalable channel counts and kernel sizes while maintaining compatibility with commercial FPGA systems through its low-voltage driving scheme and long low-loss OTDLs. As a result, our work experimentally validates that integrated photonic processors can effectively reduce the dimensions of FC layers and bridge the gap between PNNs and practical machine learning applications.

## 2. Design Principle and Device Characterization

Figure 1 illustrates the architecture and operating principles of our OCPs. The design incorporates three functionally coupled modules on a single PIC: (1) a high-bandwidth data encoding module implemented through a high-speed electro-optic modulator array, (2) a precisely calibrated delay module comprising low-loss OTDLs with arithmetic progression in path lengths, and (3) a reconfigurable kernel weight module utilizing an array of 1×2 electro-optic Mach-Zehnder interferometers (MZIs). To demonstrate the operational paradigm, we choose the following example in which an $L$-channel $M \times M$ convolutional processor executes spatial convolution on an $N \times N$ input image matrix. The processing begins by generating $M$ copies of the $N \times N$ input image, with the $i$-th copy ($i \in [1, M]$) containing rows from $i$ to $N-M+i$. Each sliced image is flattened into a one-dimensional sequence and converted into high-frequency analog voltage signals via digital-to-analog convertors (DACs). These analog signals subsequently drive the RF electrodes of respective high-speed electro-optic modulators in the data module, where the propagating lights undergo intensity modulation, effectively encoding the image data onto the laser beams. This step achieves longitudinal alignment of the original image, establishing temporal correspondence between vertically adjacent pixels. Each modulated laser is then split into $M$ paths and fed into the delay module, which provides systematically increasing delays from $0$ to $(M-1)\Delta T$ (where $\Delta T$ represents the sampling interval of the DACs) through precisely controlled optical path length differences. The $\Delta T$-spaced delays are achieved by designing long OTDLs with length increments of $(c/n_g)\Delta T$, where $c$ is the speed of light in vacuum and $n_g$ is the group index of the waveguide mode. The optical beam from each OTDL output is further split into $L$ paths and directed to the 1×2 Mach-Zehnder interferometers (MZIs) in the weight module. For the optical signal originating from the $i$-th high-speed modulator, delayed by the *j*-th OTDL, and routed to the *k*-th 1×2 MZI, the MZI imposes a weight value $W_{ij}^k$, which denotes the weight at row *i*, column *j* of the *k*-th channel convolution kernel. Due to the non-negative nature of optical intensity modulation, we employ a balanced detection scheme: the photocurrents from both output ports of each 1×2 MZI are converted via photodiodes and subsequently subtracted in the analog domain. This approach not only enables the realization of both positive and negative weight values but also reduces signal noise through balanced detection. The convolution result for each channel is obtained by summing up all corresponding photocurrents within that channel. As the optical signals propagate through the photonic circuit, the convolutional kernel automatically scans across the input image in a row-wise manner, performing continuous parallel computation.

Figure 2 (a) illustrates the layout of the fabricated 4 channel 4×4 OCP. The design incorporates 4 high-frequency modulators, 16 OTDLs, and 64 1×2 MZIs onto a LNOI chip of a footprint 3 cm × 4 cm. Each modulator consists a pair of 1×2 MMIs with tandem phase shifters - a high-frequency section for data encoding and a low-frequency section for operating point tuning. The arm lengths of the high-frequency phase shifters are set to 3 cm to achieve a sufficiently low half-wave voltage ($V_\pi$), allowing them to be driven by CMOS-compatible voltages. The 16 OTDLs possess four distinct lengths corresponding to delays of 0 ps, 500 ps, 1000 ps, and 1500 ps, respectively. Figure 2 (b) presents a photograph of the fabricated sample, captured using a digital camera and packaged with a PCB via wire bonding. The sample's high-frequency electrodes are connected to coaxial transmission lines through the PCB, with one termination linked to a high-speed analog-to-digital converter (ADC) for input signal encoding while the opposite termination interfaces with 50-ohm matched loads to maintain impedance maching and thereby minimizing signal reflections. The bias phase shifters of the four modulators and the control electrodes of the 64 sets of 1×2 MZIs are connected to an external voltage control unit via probe sets for synchronized control. Optical signals are first coupled into the photonic circuit through the right side fiber array, and propagate through the integrated device architecture for parallel processing, and finally are collected by the output fiber array on the left side.

To systematically characterize the foundational components of our devices, we performed sequential performance evaluations of the three core functional modules. The modulators within the data encoding module were characterized by applying a 1 MHz sawtooth waveform to their high-frequency RF electrodes while applying a DC bias voltage through the low-frequency operating point control electrodes. Figure 3(a) displays the resulting intensity versus applied voltage curve, showing a $V_\pi$ of 1.15 V that meets CMOS voltage driving requirements, with the modulator properly biased at its linear operating point. Figure 3(b) further demonstrates the high-speed performance of this modulator through clear, well-defined eye diagrams obtained under 2 GBaud PAM-4 modulation, confirming its excellent signal fidelity and robust multi-level signaling capability. Figure 3(c) systematically compares the ground truth values of 1000 random input signals with their corresponding output voltages after completing the full photonic processing chain (optical encoding by the modulator, photoelectric conversion, and transimpedance amplification), along with their error distribution. The statistical analysis in Figure 3(d) demonstrates that the computational deviations follow a Gaussian distribution with a standard deviation of 0.74%, corresponding to 7-bit computational precision. To characterize the delay module, we processed a serialized fashion-MNIST dataset image through the 1×8 photonic processor's data encoding module. Figure 3(e) displays the output signals from different OTDL channels, with Figure 3(f) providing an expanded temporal view. The measured delays demonstrate 500 ± 10 ps precision between adjacent channels as shown in Figure 3(g), confirming a sub-2% variation in the photonic delay line's temporal control accuracy. Finally, Figure 3(h) characterizes the weight module's performance by presenting the measured modulation curves from both output ports of a representative 1×2 MZI, demonstrating a high extinction ratio of 25 dB. This result confirms the MZI's capability for precise optical weighting operations in our devices. We further evaluated the device's fundamental convolution capabilities using standard spatial filters including edge detection and averaging kernels. The complete test setup and corresponding processing results are detailed in the Supplementary Materials. (see Section S3 and S4, Supplementary Materials)

## 3. Convolutional Neural Network Implementation

While basic operations such as edge detection and averaging kernels effectively demonstrate a convolutional processor's basic functionality, the true potential emerges when employing trainable, reconfigurable kernels optimized for specific tasks. Such adaptive kernels enable superior feature extraction while significantly reducing subsequent neural network computational overhead. In this work, we leverage our OCPs to demonstrate classification tasks across three benchmark datasets: MNIST (handwritten digits), Fashion-MNIST (apparel), and AG News (text categorization).

### 3.1. Classification of MNIST and Fashion-MNIST Datasets

Both MNIST and Fashion-MNIST datasets consist of 28×28 pixel grayscale images spanning ten categories, permitting identical convolutional neural network architectures as depicted in Fig. 4 (a). The zero-padded input images are processed through four distinct 4×4 convolutional kernels in our computational simulation, generating four 28×28 feature maps where each pixel aggregates localized 4×4 receptive field information. The optimized kernels obtained through training ensure preservation of classification-critical features after 4×4 max-pooling downsampling, ultimately condensing the original image information into four 7×7 feature maps. These maps are flattened into a 1×196 vector and subsequently mapped to a 1×10 output vector via a FC layer. This processing pipeline effectively compresses the image dimensionality from 28×28=784 to 4×7×7=196 while retaining semantically salient features, significantly reducing subsequent FC layer computational overhead without compromising accuracy. Figure 4(d) compares simulated accuracy curves for MNIST classification under two paradigms: our 4-channel 4×4 convolutional approach followed by 4×4 pooling, versus direct 2×2 pooling (both achieving 4× compression of the original image) before the 196×10 FC layer. The convolutional simulation achieves 97.53% accuracy after 20 epochs, markedly outperforming the 91.71% accuracy of the non-convolutional approach. Analogous simulation results for Fashion-MNIST are presented in Fig. 4(f). Both results demonstrate that convolutional processing maintains feature discriminability while dramatically reducing FC layer computational demands.

For each dataset, we configured the trained 4×4 convolutional kernels into our fabricated 4-channel photonic processor. From the respective test sets, 100 samples were randomly selected and encoded into the device through the data module for optical convolutional processing. The photonic computation results were subsequently combined with ReLU nonlinear activation functions implemented in software, followed by the final fully-connected layer operations. Comparative analysis between experimental and theoretical results, as shown in the confusion matrices of Figs. 4(e) and 4(g), reveals approximately 1% accuracy degradation in the hardware implementation. This marginal performance difference primarily stems from noise introduced during optoelectronic conversion processes, including photodetection noise and analog-to-digital conversion artifacts. Figure 4(b) presents the measured voltage signals from our optical convolutional processor executing four distinct convolution operations on representative samples from each dataset, along with their corresponding ground-truth theoretical values. Figure 4(c) compares the final output vectors obtained experimentally with their ideal computational counterparts. The results demonstrate that the photonic implementation introduces additional noise in the convolved features, leading to slightly higher error rates compared to numerical simulations - a finding consistent with the accuracy degradation observed in the confusion matrices.

Our experimental results demonstrate successful feature extraction on both MNIST and Fashion-MNIST datasets using the photonic processor, achieving a 4× reduction in subsequent computational overhead while maintaining comparable classification accuracy with only marginal degradation attributable to system noise.

### 3.2. Classification of AG News Dataset

In natural language processing applications, kernel size selection is particularly crucial as it determines the receptive field over text sequences, directly influencing the model's ability to capture n-gram patterns and long-range dependencies. The AG News dataset is a fundamental natural language processing benchmark, comprising 120,000 news articles evenly distributed across four categories (World, Sports, Business, Sci/Tech). Its balanced, domain-specific text makes it indispensable for evaluating text classification models, from traditional methods to deep learning systems. The AG News classification pipeline processes text through sequential dimensionality reduction stages as illustrated in Fig. 5(a). The preprocessing pipeline converts raw text into 4D word embeddings (4×200 matrix, 200=max sequence length), with positive-definite constraints applied during embedding layer training to accommodate the data encoder's non-negative value limitation. Four 4×2, two 4×4, and one 4×8 convolutional kernels then slide across this matrix, generating four 4×199, two 4×197, and one 4×193 feature maps. Subsequent 4×8 max-pooling condenses these features into seven 1×25 maps, halving the sequence length while preserving salient n-gram information. This processing compresses the original 4×200=800 input into 1×25=25 total features, a reduction of approximately 4.5×, before final mapping to the 4-class output via an FC layer with SoftMax activation.

Figure 5(b) presents the training accuracy curves for various combinations of convolutional kernel sizes, showing a consistent improvement in performance as the kernel size increases. This enhancement is due to the larger kernels' ability to capture long-range contextual patterns more effectively. We implemented these optical convolution operations by aggregating multiple operational cycles of an 1×8 OCP (Details of the device can be found in Section S2, Supplementary Materials) to reconstruct full convolution outputs. Figure 5(c) compares the measured voltage signals with the ground truth for a randomly selected sample processed using different kernels. Additionally, Figure 5(d) compares the output vector obtained from experimental results with theoretical predictions. A statistical analysis conducted on 100 randomly selected test samples (shown in Figure 5e) indicates a 3% decrease in accuracy for the fabricated OCP compared to theoretical expectations. This discrepancy is notably larger than what was observed with the MNIST and Fashion-MNIST datasets, most likely due to the lower dimensionality of word embeddings, which results in the reduced error tolerance for textual feature extraction. Scaling up to multi-channel devices in the future could significantly alleviate this limitation by allowing for the processing of higher-dimensional embeddings.

## 4. Discussion

Table 1 benchmarks the performance of representative integrated optical convolution processors, demonstrating our device's unique ability to dramatically compress subsequent FC-layer computations through large kernel sizes and independently reconfigurable high-precision weight tuning. The present 0.128 TOPS computational power is constrained by the 2 Gbaud and low-voltage AWGs required for FPGA compatibility, represents a current trade-off for system integrability. Theoretical modeling demonstrates that scaled architectures employing 16-channel 10×10 convolutional kernels with global max-pooling can reduce the FC layer dimensionality to 16×10 for MNIST dataset while preserving 97% classification accuracy, and promise computational throughput exceeding 10 TOPS while maintaining full FPGA compatibility. (see Section S5, Supplementary Materials) This compact FC layer size is fully compatible with current PIC implementations, providing strong motivation for future scaling up of OCPs. Another critical consideration is that our current implementation still relies on off-chip analog photocurrent summation for convolutional operations. Future research will explore three primary approaches to overcome this limitation: (1) The monolithic integration of high-quality photodetectors on TFLN platforms, which is essential for all TFLN-based optical computing architectures and has demonstrated significant experimental progress in recent years, confirming its technical viability;[38–41] (2) The development of optimized architectures utilizing highly compact, low-loss waveguide crossings and combiners on TFLN to enable optical-domain addition, which would significantly reduce dependence on photodetection; (3) Strategic incorporation of monolithic integration of WDM modules to further enhance parallel processing capability. An additional consideration is that while most architectures successfully implement negative-valued convolutional weights through balanced detection, their requirement for positive-definite input data fundamentally constrains model performance. Future research will investigate scalable optical computing schemes that accommodate negative values for both input data and weight values.

## 5. Conclusion

To conclude, we demonstrate monolithically integrated lithium niobate OCPs that significantly reduce the computational overhead of FC layers while maintaining competitive classification accuracy across image and text recognition tasks. By leveraging large, programmable photonic convolution kernels, our architecture achieves a 4× reduction in FC layer dimensions for image processing (from 784×10 to 196×10) and a 4.5× reduction for text classification (from 800×4 to 175×4), with accuracies of 96% (MNIST), 86% (Fashion-MNIST), and 84.6% (AG News). This work bridges the gap between photonic neural networks inherently suffering from low dimension numbers and practical machine learning tasks relying a large number of dimensions. Future up-scaling to higher channel counts and larger kernel dimensions promises to unlock >10 TOPS performance while preserving FPGA compatibility, establishing a pathway toward ultra-efficient photonic accelerators for next-generation artificial intelligence.

**Acknowledgements**

R.L., R.W., and Y. Z. contributed equally to this work. The work was supported by the National Natural Science Foundation of China (12192251, 12334014, 12134001, 12304418, 12474378). Innovation Program for Quantum Science and Technology (2021ZD0301403).Shanghai Municipal Science and Technology Major Project (2019SHZDZX01).

**Data Availability Statement**

The data that support the findings of this study are available on request from the corresponding author. The data are not publicly available due to privacy or ethical restrictions.

**References**

[1] M. M. Waldrop, *Nature News* **2016**, *530*, 144.

[2] Y. LeCun, Y. Bengio, G. Hinton, *Nature* **2015**, *521*, 436.

[3] Y. Shen, N. C. Harris, S. Skirlo, M. Prabhu, T. Baehr-Jones, M. Hochberg, X. Sun, S. Zhao, H. Larochelle, D. Englund, M. Soljačić, *Nature Photon* **2017**, *11*, 441.

[4] S. Pai, Z. Sun, T. W. Hughes, T. Park, B. Bartlett, I. A. D. Williamson, M. Minkov, M. Milanizadeh, N. Abebe, F. Morichetti, A. Melloni, S. Fan, O. Solgaard, D. A. B. Miller, *Science* **2023**, *380*, 398.

[5] Z. Li, Z. Deng, J. Liu, C. Bian, J. Li, Z. Ruan, R. Gan, Z. Chen, K. Chen, C. Guo, L. Liu, S. Yu, *Laser & Photonics Reviews* **2025**, *19*, 2402016.

[6] S. Bandyopadhyay, A. Sludds, S. Krastanov, R. Hamerly, N. Harris, D. Bunandar, M. Streshinsky, M. Hochberg, D. Englund, *Nat. Photon.* **2024**, *18*, 1335.

[7] Z. Xu, T. Zhou, M. Ma, C. Deng, Q. Dai, L. Fang, *Science* **2024**, *384*, 202.

[8] S. Hong, J. Wu, Y. Xie, X. Ke, H. Li, L. Lyv, Y. Peng, Q. Yao, Y. Shi, K. Wang, L. Zhuang, P. Wang, D. Dai, *Nat Commun* **2025**, *16*, 288.

[9] Z. Lin, B. J. Shastri, S. Yu, J. Song, Y. Zhu, A. Safarnejadian, W. Cai, Y. Lin, W. Ke, M. Hammood, T. Wang, M. Xu, Z. Zheng, M. Al-Qadasi, O. Esmaeeli, M. Rahim, G. Pakulski, J. Schmid, P. Barrios, W. Jiang, H. Morison, M. Mitchell, X. Guan, N. A. F. Jaeger, L. A. Rusch, S. Shekhar, W. Shi, S. Yu, X. Cai, L. Chrostowski, *Nat Commun* **2024**, *15*, 9081.

[10] Y. Zheng, R. Wu, Y. Ren, R. Bao, J. Liu, Y. Ma, M. Wang, Y. Cheng, *Laser & Photonics Reviews* **2024**, *18*, 2470060.

[11] Z. Hu, S. Li, R. L. T. Schwartz, M. Solyanik-Gorgone, M. Miscuglio, P. Gupta, V. J. Sorger, *Laser & Photonics Reviews* **2022**, *16*, 2200213.

[12] W. Shi, Z. Huang, H. Huang, C. Hu, M. Chen, S. Yang, H. Chen, *Light Sci Appl* **2022**, *11*, 121.

[13] H. H. Zhu, J. Zou, H. Zhang, Y. Z. Shi, S. B. Luo, N. Wang, H. Cai, L. X. Wan, B. Wang, X. D. Jiang, J. Thompson, X. S. Luo, X. H. Zhou, L. M. Xiao, W. Huang, L. Patrick, M. Gu, L. C. Kwek, A. Q. Liu, *Nat Commun* **2022**, *13*, 1044.

[14] C. Feng, J. Gu, H. Zhu, Z. Ying, Z. Zhao, D. Z. Pan, R. T. Chen, *ACS Photonics* **2022**, *9*, 3906.

[15] Z. Chen, A. Sludds, R. Davis, I. Christen, L. Bernstein, L. Ateshian, T. Heuser, N. Heermeier, J. A. Lott, S. Reitzenstein, R. Hamerly, D. Englund, *Nat. Photon.* **2023**, *17*, 723.

[16] S. Xu, J. Wang, H. Shu, Z. Zhang, S. Yi, B. Bai, X. Wang, J. Liu, W. Zou, *Light Sci Appl* **2021**, *10*, 221.

[17] S. Hua, E. Divita, S. Yu, B. Peng, C. Roques-Carmes, Z. Su, Z. Chen, Y. Bai, J. Zou, Y. Zhu, Y. Xu, C. Lu, Y. Di, H. Chen, L. Jiang, L. Wang, L. Ou, C. Zhang, J. Chen, W. Zhang, H. Zhu, W. Kuang, L. Wang, H. Meng, M. Steinman, Y. Shen, *Nature* **2025**, *640*, 361.

[18] L. Alzubaidi, J. Zhang, A. J. Humaidi, A. Al-Dujaili, Y. Duan, O. Al-Shamma, J. Santamaría, M. A. Fadhel, M. Al-Amidie, L. Farhan, *J Big Data* **2021**, *8*.

[19] A. Krizhevsky, I. Sutskever, G. E. Hinton, *Commun. ACM* **2017**, *60*, 84.

[20] M. D. Zeiler, R. Fergus, In *Lecture Notes in Computer Science*, **2014**, p. 818.

[21] J. Feldmann, N. Youngblood, M. Karpov, H. Gehring, X. Li, M. Stappers, M. Le Gallo, X. Fu, A. Lukashchuk, A. S. Raja, J. Liu, C. D. Wright, A. Sebastian, T. J. Kippenberg, W. H. P. Pernice, H. Bhaskaran, *Nature* **2021**, *589*, 52.

[22] X. Yu, Z. Wei, F. Sha, X. Wang, Y. Chu, Z. Wang, X. Han, H. Wang, S. Yi, Y. Cheng, G. Hu, P. Xie, *eLight* **2025**, *5*, 10.

[23] S. Xu, J. Wang, S. Yi, W. Zou, *Nat Commun* **2022**, *13*, 7970.

[24] B. Bai, Q. Yang, H. Shu, L. Chang, F. Yang, B. Shen, Z. Tao, J. Wang, S. Xu, W. Xie, W. Zou, W. Hu, J. E. Bowers, X. Wang, *Nat Commun* **2023**, *14*, 66.

[25] X. Xu, M. Tan, B. Corcoran, J. Wu, A. Boes, T. G. Nguyen, S. T. Chu, B. E. Little, D. G. Hicks, R. Morandotti, A. Mitchell, D. J. Moss, *Nature* **2021**, *589*, 44.

[26] X. Meng, G. Zhang, N. Shi, G. Li, J. Azaña, J. Capmany, J. Yao, Y. Shen, W. Li, N. Zhu, M. Li, *Nat Commun* **2023**, *14*, 3000.

[27] X. Zhang, Z. Sun, Y. Zhang, J. Shen, Y. Chen, M. Sun, C. Shu, C. Zeng, Y. Jiang, Y. Tian, J. Xia, Y. Su, *Laser & Photonics Reviews* **2025**, *19*, 2401583.

[28] Y. Xie, X. Ke, S. Hong, Y. Sun, L. Song, H. Li, P. Wang, D. Dai, *Science Advances* **2025**, *11*, eads7475.

[29] N. C. Harris, J. Carolan, D. Bunandar, M. Prabhu, M. Hochberg, T. Baehr-Jones, M. L. Fanto, A. M. Smith, C. C. Tison, P. M. Alsing, D. Englund, *Optica, OPTICA* **2018**, *5*, 1623.

[30] A. Ribeiro, A. Ruocco, L. Vanacker, W. Bogaerts, *Optica, OPTICA* **2016**, *3*, 1348.

[31] H. Zhang, M. Gu, X. D. Jiang, J. Thompson, H. Cai, S. Paesani, R. Santagati, A. Laing, Y. Zhang, M. H. Yung, Y. Z. Shi, F. K. Muhammad, G. Q. Lo, X. S. Luo, B. Dong, D. L. Kwong, L. C. Kwek, A. Q. Liu, *Nat Commun* **2021**, *12*, 457.

[32] J. Feldmann, N. Youngblood, C. D. Wright, H. Bhaskaran, W. H. P. Pernice, *Nature* **2019**, *569*, 208.

[33] J. He, J. Qiang, Y. Dong, J. Wang, T. Dong, G. Yue, R. Zhuang, M. Lv, S. Yu, Z. Lin, X. Cai, Y. Yang, G. Wu, Y. Li, **2025**, DOI 10.48550/arXiv.2506.18310.

[34] Y. Ren, B.-Y. Nan, R.-B. Wu, Y. Zheng, R.-X. Liu, X.-W. Wang, Y.-P. Song, M. Wang, Y. Cheng, *Chinese Phys. Lett.* **2025**, *42*, 070401.

[35] R. Wu, M. Wang, J. Xu, J. Qi, W. Chu, Z. Fang, J. Zhang, J. Zhou, L. Qiao, Z. Chai, J. Lin, Y. Cheng, *Nanomaterials* **2018**, *8*, 910.

[36] Y. Zheng, H. Zhong, H. Zhang, L. Song, J. Liu, Y. Liang, Z. Liu, J. Chen, J. Zhou, Z. Fang, M. Wang, L. Li, R. Wu, Y. Cheng, *Phys. Rev. Res.* **2023**, *5*, 033206.

[37] L. Song, J. Chen, R. Wu, Y. Zheng, Z. Liu, G. Wang, C. Sun, M. Wang, Y. Cheng, *Opt. Lett., OL* **2023**, *48*, 2261.

[38] Q. Dong, X. Sun, L. Gao, Y. Zheng, R. Wu, Y. Cheng, *Nanomaterials* **2025**, *15*, 72.

[39] C. Jin, C. Wang, L. Qu, D. Zhang, Q. Liu, J. You, D. Zheng, W. Wu, W. Cai, M. Ren, J. Xu, *Laser & Photonics Reviews* **2023**, *17*, 2300503.

[40] K. Xia, H. Liu, Y. Qiu, S. Zheng, Y. Dan, Q. Zhong, Y. Dong, X. Zhao, T. Hu, *Opt. Lett., OL* **2024**, *49*, 3162.

[41] Y. Guo, Y. Qiu, S. Zheng, Q. Zhong, Y. Dong, X. Zhao, T. Hu, *Appl. Opt., AO* **2025**, *64*, 5115.

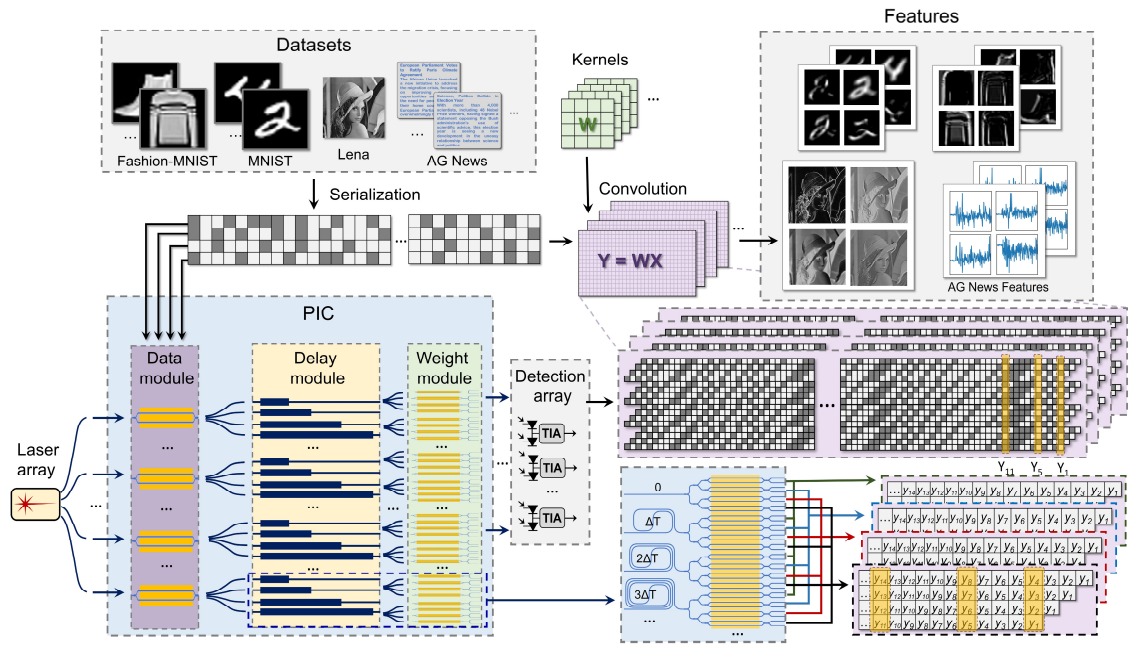

**Figure 1.** Schematic of the integrated photonic convolutional processor featuring monolithic data loading (electro-optic modulators), delay (OTDL array), and weighting (MZI bank) modules that enable parallel optical computation through $\Delta T$-spaced delays and balanced photodetection.

**Table 1.** Benchmarking integrated optical convolution processors

| Platform and schema | | Channel count | Kernel size | Weights freedom | Off-chip components | Precision | Modulation baud rate (Gbaud) | Compute speed (Tops s$^{-1}$) | Required FC layer dimensions for MNIST | Accuracy on MNIST |
|---|---|---|---|---|---|---|---|---|---|---|
| Si | MRR-VMM[23] | 4 | 1×3 | 12 | Modulators, | 7-bit | 20 | 0.24 | / | 97.90% (KTH dataset) |
| | SiN comb PCM-VMM[21] | 4 | 3×3 | 36 | WDM, Modulators | 7-bit | 2 | 0.288 | / | 95.30% |
| | AlGaAs comb MRR-VMM[24] | 1 | 2×2 | 4 | WDM | 5-bit | 17 | 0.068 | 2187×512×10 | 96.60% |
| | MZI mesh and OTDLs[28] | 4 | 2×2 | 16 | Modulators | 6-bit | 40 | 1.28 | 676×128×10 | 97% |
| SiN | MMI -VMM OCPU[26] | 4 | 2×2 | 4 | WDM, OTDLs, Modulators | 5-bit | 16.6 | 0.27 | 1568×10 | 92.17% |
| | SiN comb TOPs-CA[25] | 3 | 5×5 | / | Not integrated | 7-bit | 62.9 | 5.66 | / | 88.00% |
| TFLN | M-DPCA[27] | 4 | 2×2 | 4 | WDM, OTDLs | 5-bit | 128 | 8.19 | 2704×320×10 | 94.9% |
| | EO-comb[33] | 3 | 3×3 | 27 | OTDLs | 4-bit | 30 | 1.62 | 2028×256×10 | 95.0% |
| | MZI-OTDLs (this work) | 4 | 4×4 | 64 | Monolithically integrated | 7-bit | 2 | 0.128 | 196×10 | 96% |
| | MZI-OTDLs (expected from this work) | 16 | 10×10 | 1600 | Monolithically integrated | 7-bit | 2 10 | 3.2 16 | 16×10 | 97% |

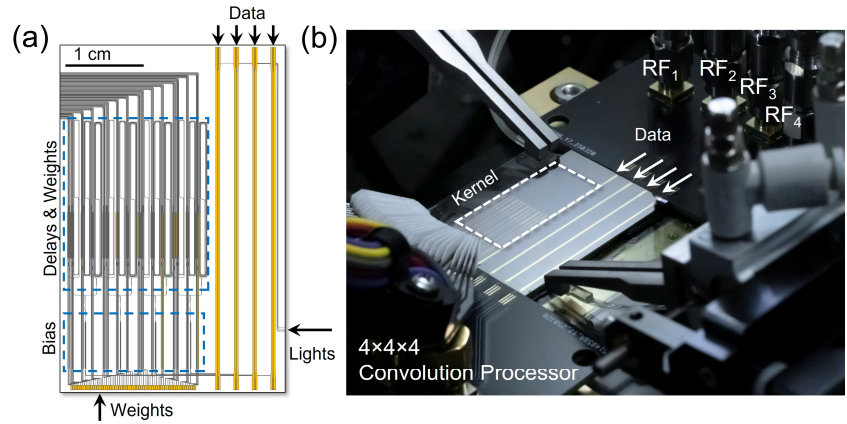

**Figure 2.** (a) Layout of the fabricated 4-channel 4×4 lithium niobate photonic processor (3×4 cm) integrating: four high-speed electro-optic modulators with 3-cm phase shifters (achieving CMOS-compatible Vπ), 16 optical true delay lines (OTDLs) providing various time delays from 0 to 1500 ps seperated by 500 ps increment, and 64 reconfigurable 1×2 Mach-Zehnder interferometers (MZIs) for weight implementation. (b) Photograph of the packaged device showing wire-bonded connections to control electronics, with high-frequency electrodes interfacing through PCB-mounted coaxial lines to ADCs.

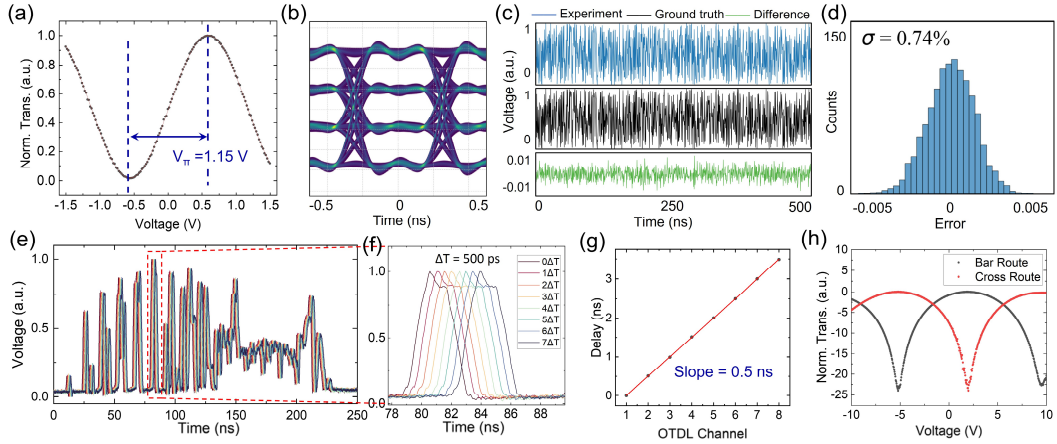

**Figure 3.** (a) Measured transfer curve of the electro-optic modulator showing a CMOS-compatible Vπ of 1.15 V under 1 MHz sawtooth modulation. (b) Clear 2 GBaud PAM-4 eye diagrams demonstrating robust multi-level signaling. (c) Comparison of 1000 input/output signals through the full photonic processing chain, with (d) Gaussian-distributed errors (0.32% STD, 8-bit precision). (e) OTDL output signals from MNIST processing, with (f) expanded temporal view and (g) confirmed 500±10 ps inter-channel delays (sub-2% error). (h) MZI modulation curves showing 25 dB extinction ratio for precise optical weighting.

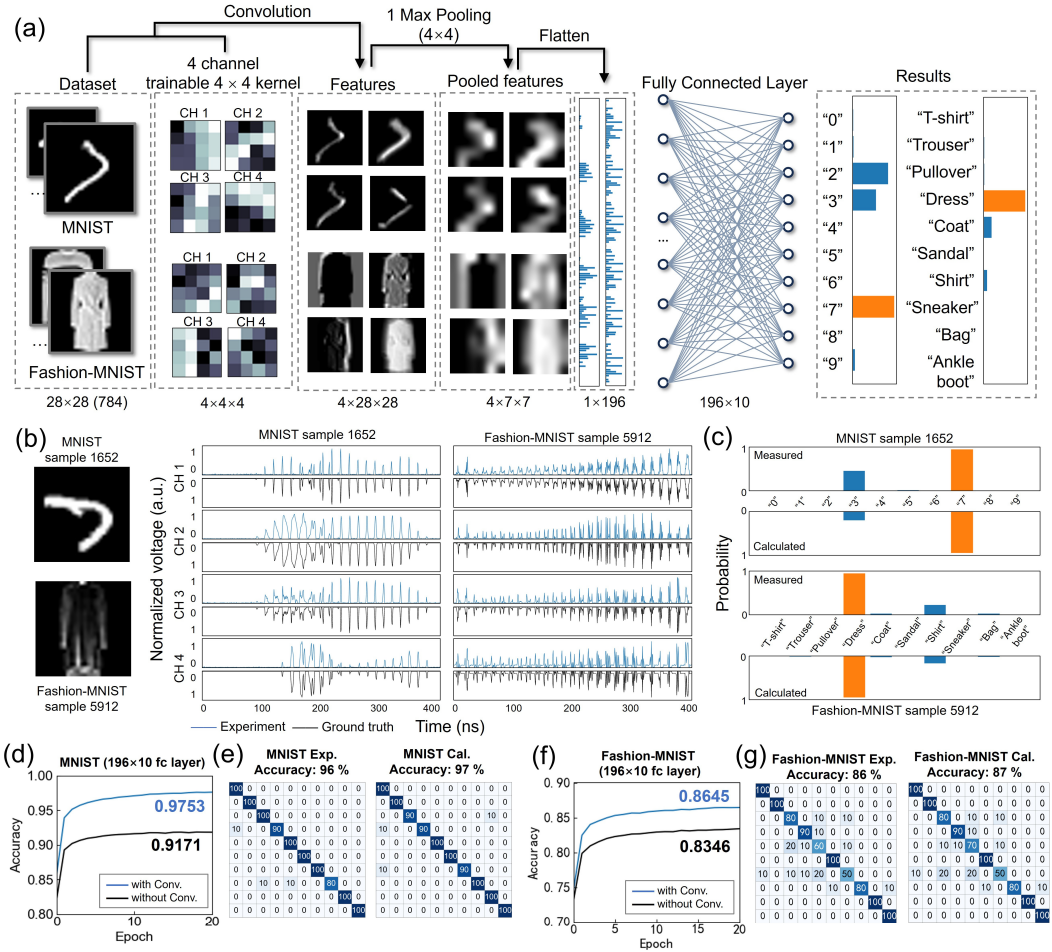

**Figure 4.** (a) Network architecture for MNIST/Fashion-MNIST classification using four 4×4 convolutional kernels, max-pooling, and fully-connected layers. (b) Measured voltage signals of two samples from MNIST/Fashion-MNIST dataset after optical convolution operations versus theoretical ground truth. (c) Experimental versus simulated output vectors of the two samples. (d) Simulated accuracy curves for MNIST classification under two paradigms: our 4-channel 4×4 convolutional approach followed by 4×4 pooling, versus direct 2×2 pooling (both achieving 4× compression) before the 196×10 FC layer. (e) Confusion matrices of the experimental and theoretical results of MNIST clasification. (f) Corresponding Fashion-MNIST training accuracy curves. (g) Confusion matrices of the experimental and theoretical results of Fashion-MNIST clasification.

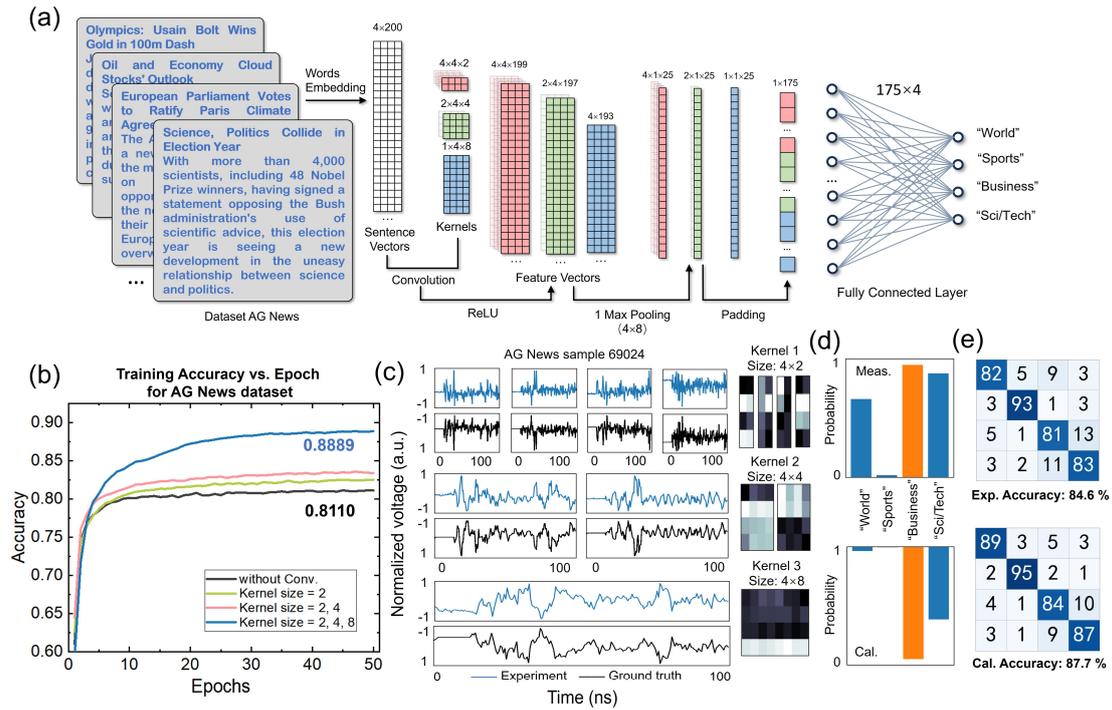

**Figure 5.** (a) AG News processing pipeline: 4D word embeddings transform input text into a 4×200 matrix, processed by multi-scale kernels (4×2/4×4/4×8) to generate feature maps, then condensed via 4×8 max-pooling into seven 1×25 features (4.5× dimensionality reduction) before FC classification. (b) Training accuracy curves for different kernel sizes, showing improved performance with larger kernels due to enhanced long-range context capture. (c) Measured vs. ground-truth voltage signals for sample processed with different kernels. (d) Experimental vs. theoretical output vectors. (e) Confusion matrix for 100 test samples showing 3% accuracy drop in fabricated OCP versus theoretical expectations, attributed to low embedding dimensionality.

# Supporting Information

## Monolithically Integrated Optical Convolutional Processors on Thin Film Lithium Niobate

*Ruixue Liu, Rongbo Wu\*, Yong Zheng, Yuan Ren, Boyang Nan, Min Wang\*, and Ya Cheng\**

**1. Device Fabrication**

Our device is fabricated using a home-developed process. This process employs femtosecond laser direct writing to pattern a Cr hard mask, followed by chemo-mechanical polishing (CMP) to transfer the mask patterns into the underlying thin-film lithium niobate (TFLN). The TFLN structure comprises a 500-nm-thick TFLN layer bonded to a buried $SiO_2$ layer on a 500-μm-thick silicon substrate (NANOLN). The fabrication process flows the following steps as shown in Figure S1: (1) Deposit a 200-nm-thick chromium (Cr) film onto a commercial 4-inch thin-film lithium niobate (TFLN) wafer using magnetron sputtering. (2) Pattern pre-designed photonic structures on the Cr film using femtosecond laser direct writing system. (3) Transfer the patterns onto the TFLN layer through chemo-mechanical polishing (CMP). This enables waveguides with ultra-low sidewall roughness, yielding exceptionally low propagation loss of about 0.03 dB/cm. (4) Remove residual Cr using chromium etchant. (5) Deposit a 1.5 μm thick layer of $SiO_2$ on the wafer as the waveguide cladding layer. (6) Depositing a chromium-gold-titanium (Cr-Au-Ti) layer on the wafer with a thickness of 10 nm, 500 nm, and 100 nm by magnetron sputtering. (7) Pattern the electrodes masks onto the Ti layer for gold electrodes using femtosecond laser lithography. (8) Wet-etch non-electrode regions of Au using Ti as the etch mask. (9) Pattern the underlying Cr layer via femtosecond laser direct writing. (10) Remove residual Ti masks by wet etching to finalize the electrode structures.

**2. Photograph of the single channel 1×8 OCP**

Figure S2 (a) illustrates the layout of the fabricated 8-channel lithium niobate photonic processor (1×4 cm). The design integrates: (1) a high-speed electro-optic modulator with a 3 cm phase shifter achieving CMOS-compatible $V_\pi$, (2) eight optical true delay lines (OTDLs) providing programmable delays from 0 to 3500 ps in 500 ps increments, and (3) reconfigurable 1×2 Mach-Zehnder interferometers (MZIs) for weight implementation. Figure S2 (b) shows a photograph of the packaged device, where wire-bond connections interface the chip with control electronics, and PCB-mounted coaxial cables link the high-frequency electrodes to an ADC. Figures S2 (c-e) present microscope images detailing three critical subsystems: (c) the 8 low-loss OTDLs with progressively increasing lengths, (d) high-frequency electrodes with precision wire bonds to the PCB, and (e) low-frequency electrodes for weight and oparating point turning.

### 3. Experimental Measurement Setup

To characterize the convolution operation capability of the fabricated sample, we employed the setup illustrated in Figure S3. Four laser beams generated by laser diodes (CTL1550, TOPTICA Photonics Inc., Farmington, New York, USA) were amplified using an erbium-doped fiber amplifier (EDFA, KY-EDFA-HP-37-FA, KEYANG PHOTONICS, Beijing, China) and converted to the transverse electric (TE) mode via four in-line polarization controllers (Polarization Synthesizer, PSY 201, General Photonics Corp., Chino, California, USA). These beams were then input into the sample through a fiber array. The input data were transformed into four high-frequency voltage signals. These signals were generated by a four-channel arbitrary waveform generator (AWG 5000 SERIES, Tektronix) and fed into the high-frequency modulators within the sample. A 128-channel PXIe system (PXIe-6739, National Instruments Corp., Austin, Texas, USA) generated both the 64 weight values applied to the control electrodes of the 1×2 MZIs (for the 4-channel 4×4 convolutional kernels) and the 4 operating point tuning parameters for the high-frequency modulators. The laser signals transmitted through the sample were converted into electrical signals by photodetector arrays (KY-APRM-10G-I-SM, KEYANG PHOTONICS, Beijing, China). These electrical signals were received by a high-speed real-time oscilloscope (MSO64B, Tektronix Inc., Beaverton, Oregon, USA).

### 4. Basic Kernels Test

Figure S4 demonstrates a basic validation of our optical convolutional processor, showing the convolution of six grayscale images using two standard 4×4 kernels (edge detection and blur).

The experimental results align closely with theoretical predictions, confirming that our optical implementation introduces negligible distortion to these simple convolutional operations. This successful verification establishes a foundation for implementing more complex image processing tasks in the photonic domain.

**5. Extreme Feature Compression Using Large-Receptive-Field Convolutional Networks**

To validate that larger convolutional kernels can more effectively reduce the computational overhead of subsequent fully-connected layers, we trained a 16-channel 10×10 convolutional kernel-based network on the MNIST classification task. The neural network architecture processes MNIST's 28×28 grayscale images through 16 optimized 10×10 convolutional kernels, with each kernel generating feature maps where every pixel integrates information from an extended 10×10 receptive field. Global max-pooling subsequently reduces each 28×28 feature map to its single most salient activation, compressing the spatial dimensionality to 1×1 per channel and yielding a compact 1×16 vector representation. This achieves an extreme dimensionality reduction from 784 (28×28) to just 16 while maintaining 97% classification accuracy on the MNIST test set.

The effectiveness of this architecture stems from the careful balance between aggressive dimensionality reduction and intelligent feature preservation. While global max-pooling discards the vast majority of spatial information from each feature map, the exceptionally large 10×10 convolutional kernels ensure complete retention of the most discriminative features through their expansive receptive fields. The preservation of 97% classification accuracy despite this extreme compression provides definitive proof that the network successfully retains all critical category-defining information during this process. The training dynamics and classification performance are presented in Figure S5. Figure S5(a) shows the convergence curve during training, demonstrating the rapid achievement and stability of the 97% accuracy. Figure S5(b) presents the corresponding confusion matrix, revealing particularly strong performance on digit categories where the large 10×10 receptive fields effectively capture characteristic stroke patterns.

**6. Compensation techniques for electro-optic relaxation problem in TFLN devices**

While TFLN is a leading EO platform, its modulators suffer from temporal EO relaxation (ms to hours). This instability hinders low-frequency and long-duration applications, including optical neural networks requiring precise weight retention. A favorable phenomenon is that EO relaxation in TFLN does not occur when the electrical frequency applied to the modulator is above 1KHz. In this paper, we propose a method to address the EO relaxation problem in lithium niobate by utilizing medium-high frequency synchronous alternating positive-negative signals. Specifically, instead of applying a constant bias voltage, a 40 kHz synchronous alternating positive-negative square wave signal is applied to the modulator used for imposing weight values. The first half-cycle of the square wave signal matches the target bias voltage, while the second half-cycle voltage is the opposite of the target bias voltage. During testing, data are only collected from the first half-cycle, with data from the second half-cycle discarded. Although this method sacrifices 50% of the modulation rate, it is compatible with all devices.

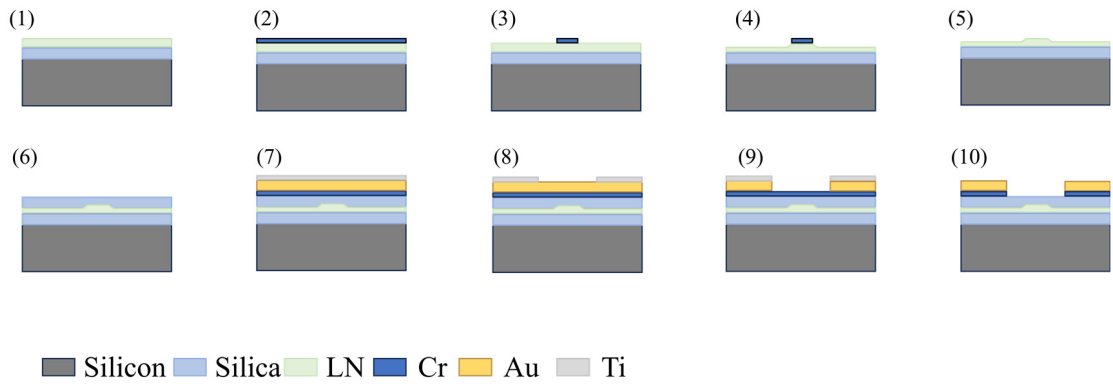

**Figure S1.** Schematic diagram of the fabrication flows of PLACE technology.

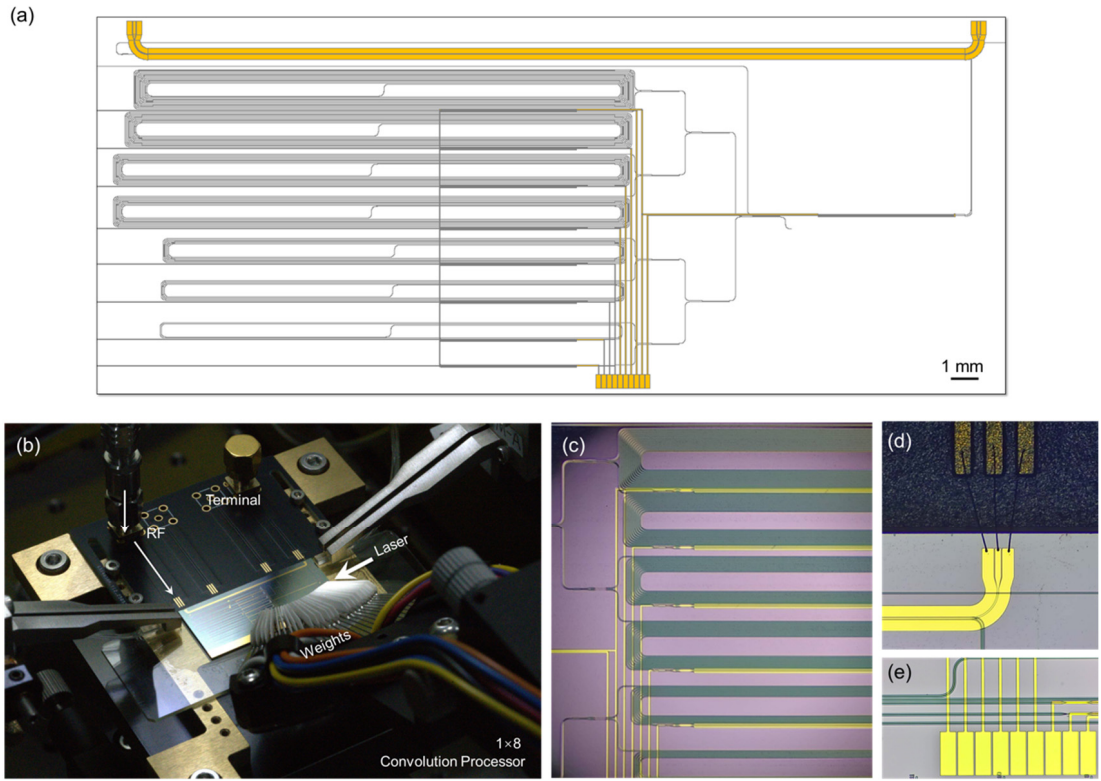

**Figure S2.** (a) Layout of the fabricated 8-channel 1×8 lithium niobate photonic processor (1×4 cm), which integrates a high-speed electro-optic modulator with a 3 cm phase shifter (achieving CMOS-compatible Vπ), 8 optical true delay lines (OTDLs) providing delays from 0 to 3500 ps in 500 ps increments, and reconfigurable 1×2 MZIs for weight implementation. (b) Photograph of the packaged device, showing wire-bond connections to the control electronics. (c–e) Microscope images of the 1×8 lithium niobate photonic processor.

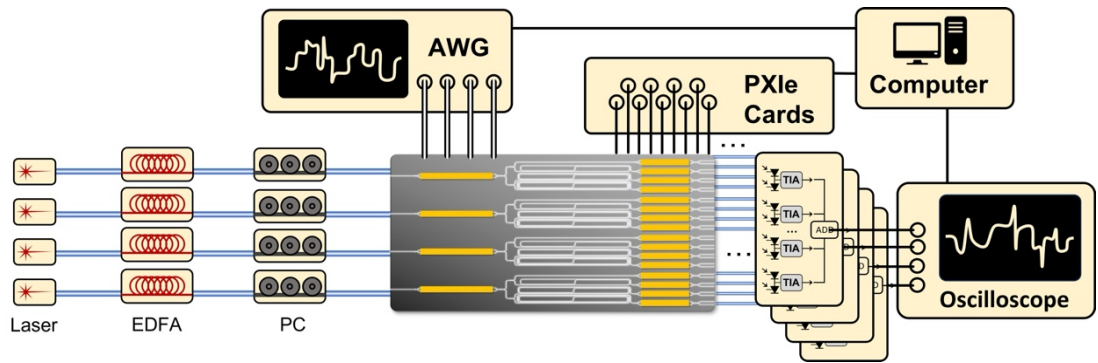

**Figure S3.** Experimental setup for optical convolution characterization.

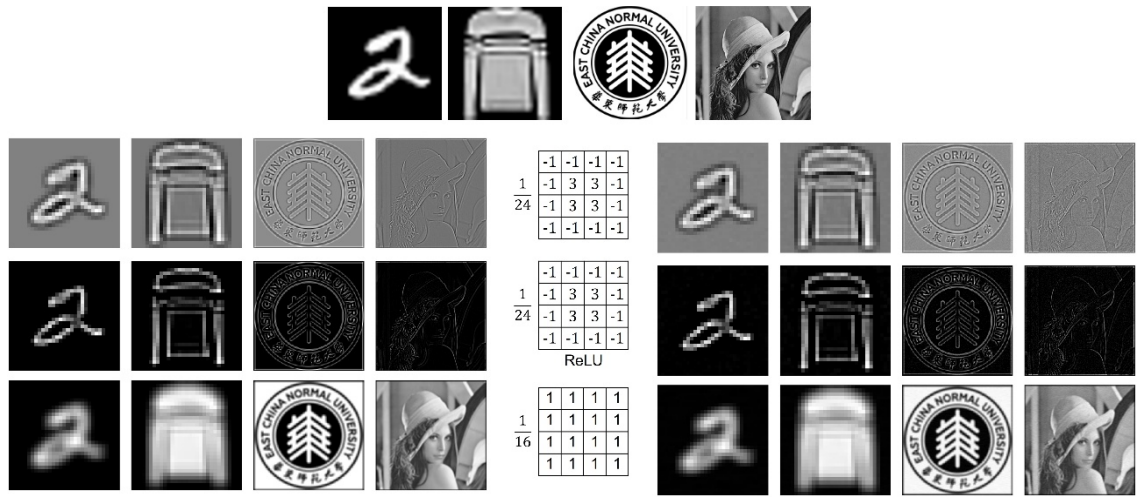

**Figure S4.** Optical convolutional processing validation. Experimental results (right) of convolving six grayscale images with standard 4×4 edge-detection and blur kernels show excellent agreement with theoretical predictions (left).

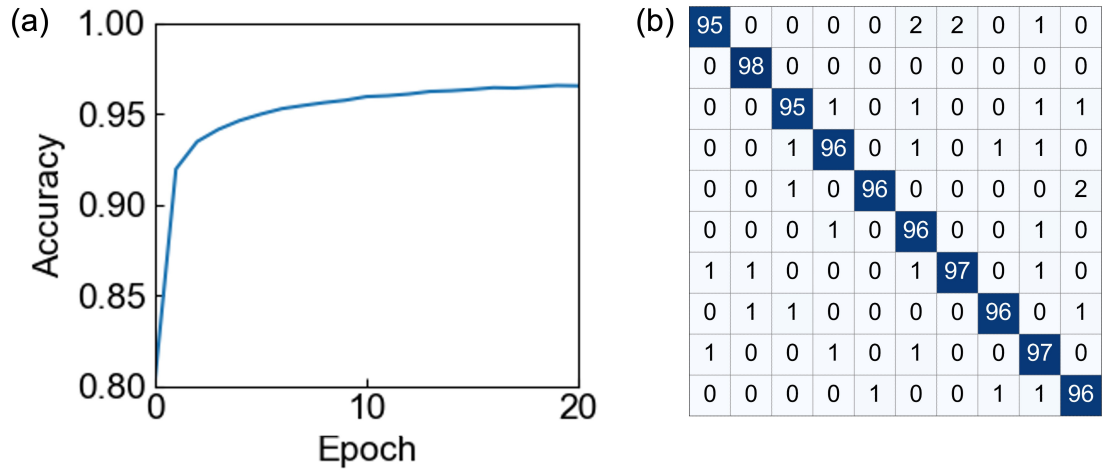

**Figure S5.** Training dynamics and classification performance of the 16-channel 10×10 convolutional network. (a) Training curve showing convergence to 97.3% test accuracy on MNIST, demonstrating stable learning despite extreme dimensionality reduction. (b) Confusion matrix revealing strong diagonal dominance.